\documentclass[a4paper,twoside,twocolumn,english,aps,prl,superscriptaddress,floatfix]{revtex4}
\usepackage{ae,aecompl}
\usepackage[T1]{fontenc}
\usepackage[latin9]{inputenc}
\usepackage{color}
\usepackage{float}
\usepackage{amsmath}
\usepackage{hyperref,amsmath}
\usepackage{graphicx,bm}
\usepackage{babel}


\begin{document}

\title{Bose-Einstein condensation in quantum glasses}

\author{Giuseppe Carleo}

\affiliation{SISSA -- Scuola Internazionale Superiore di Studi Avanzati and CNR-INFM
DEMOCRITOS -- National Simulation Center, via Beirut 2-4, I-34014
Trieste, Italy.}

\author{Marco Tarzia}

\affiliation{Laboratoire de Physique Th\'eorique de la Mati\`ere Condens\'ee, Universit\'e Pierre et Marie Curie-Paris 6, 
UMR CNRS 7600, 4 place Jussieu, 75252 Paris Cedex 05, France.}

\author{Francesco Zamponi}

\affiliation{Princeton Center for Theoretical Science, Princeton University, Princeton, New Jersey 08544, USA}
\affiliation{Laboratoire de Physique Th\'eorique, Ecole Normale Sup\'erieure, 
UMR CNRS 8549, 24 Rue Lhomond, 75231 Paris Cedex 05, France.}

\pacs{61.43.Fs; 05.30.Jp; 67.80.K-}

\begin{abstract}
The role of geometrical 
frustration in strongly interacting bosonic
systems is studied with a combined numerical and analytical approach. 
We demonstrate the existence of a novel quantum phase featuring both Bose-Einstein
condensation and spin-glass behaviour. The differences between such a phase and the 
otherwise insulating {}``Bose glasses\emph{''} are elucidated.
\end{abstract}
\maketitle

\paragraph*{Introduction. }

Quantum particles moving in a disordered environment exhibit a plethora
of non-trivial phenomena. The competition between disorder and quantum
fluctuations has been the subject of vast literature \citep{Anderson:1958sf,Fisher:1989gf}
in past years, with a renewed interest following from the exciting
frontiers opened by the experimental research with cold-atoms \citep{Roati:2008gf,Billy:2008ve}.
One of the most striking features resulting from the presence of a
disordered external potential is the appearance of localized
states~\citep{Anderson:1958sf}. Localization happens both
for fermions and bosons \citep{Fisher:1989gf},
but in the latter case one has to introduce repulsive interactions to prevent 
condensation of particles in the lowest energy state. This results in
the existence of an {\it insulating} phase called {}``Bose glass'', characterized by a finite
compressibility and gapless density excitations in sharp contrast to the Mott insulating 
phase~\cite{Fisher:1989gf,Krauth:1991db}. 

On the other hand, latest research stimulated by the discovery of a supersolid phase
of Helium has brought to the theoretical foresight of a {}``superglass''
phase \citep{boninsegni:105301,biroli:224306}, corroborated by recent experimental evidence \cite{B.Hunt05012009},
where a metastable amorphous solid
features both condensation and superfluidity, {\it in absence of any random external potential}.
The apparent irreconcilability, between the current picture of insulating ``Bose glasses" and the emergence of this novel phase of matter, 
calls for a moment of thought. 
Although it has been recently demonstrated that \emph{attractively} interacting lattice bosons can overcome the localization induced by an external random potential and feature a coexistence of superfluidity and amorphous order \cite{Dang:2009la}, a general understanding of the physics of Bose-Einstein condensation in quantum glasses and in presence of purely repulsive interactions is still in order.
In particular, we wonder what could be the possible microscopic mechanism leading to super-glassines and if the external disorder, current paradigm in the description of quantum glasses, could be replaced by some other mechanism. 

In this Letter we show that geometrical frustration is the missing ingredient. 
Geometrical frustration is a well recognised feature of disordered phases
in which the translational symmetry is not explicitly broken by any external potential.
Examples are spin liquids phases of frustrated
magnets \citep{Capriotti:2001ly}, valence-bond glasses
\citep{tarzia_biroli} and the \textit{order-by-disorder} mechanism inducing supersolidity on frustrated lattices \cite{Wessel:2005sf}. 
Another prominent manifestation
of frustration is the presence of a large number of metastable states
that constitutes the fingerprint of spin-glasses. 
When quantum fluctuations and geometrical frustration meet, their interplay 
raises nontrivial questions on the possible realisation of relevant phases of matter.  
Most pertinently to our purposes: can quantum fluctuations stabilise a superglass phase in a \textit{self-disordered} environment induced by geometrical frustration? 
Hereby we answer this question demonstrating that repulsively interacting bosons can feature a low-temperature phase characterised both by spin-glass order and Bose-Einstein condensation.
Such a frustration induced superglass sheds light onto a novel mechanism
for glass formation in bosonic systems noticeably different from the
localization effects leading to ``Bose glass" insulators and paving the way to a better understanding of 
this new phase of the matter.  

\begin{figure}
\includegraphics[width=6cm]{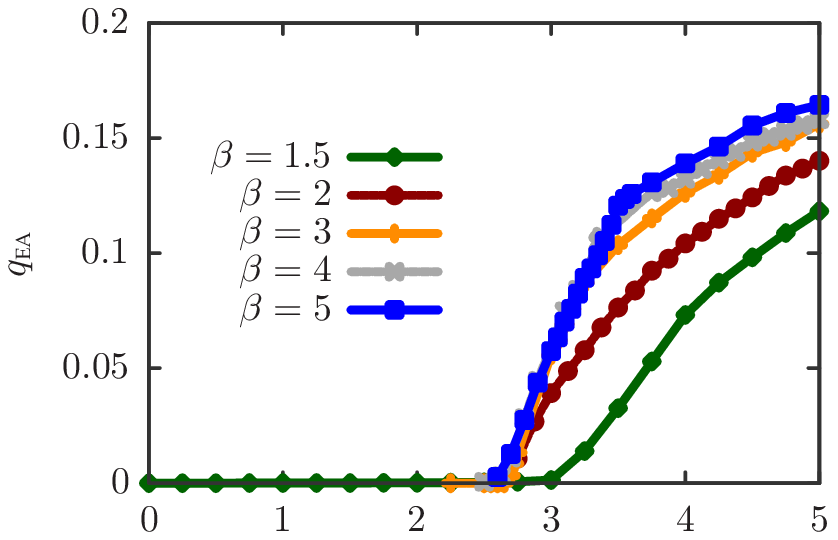}
\includegraphics[width=6cm]{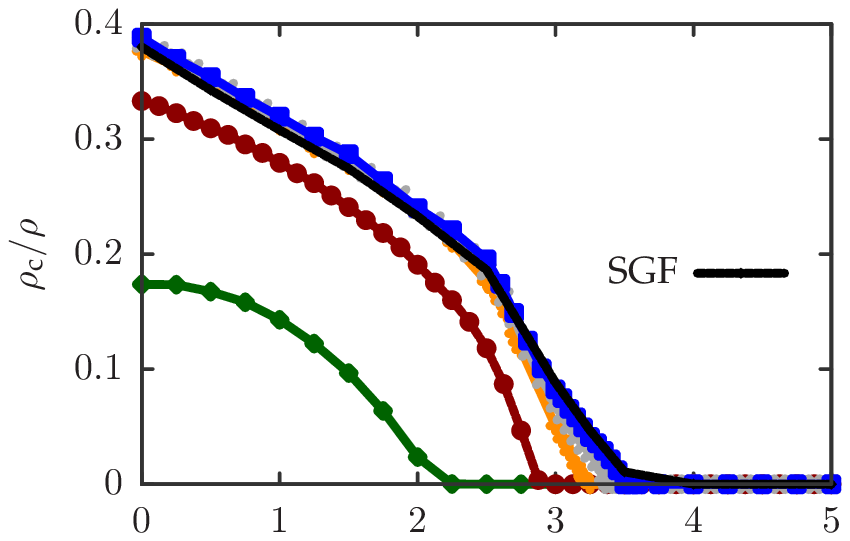}
\includegraphics[width=6.2cm]{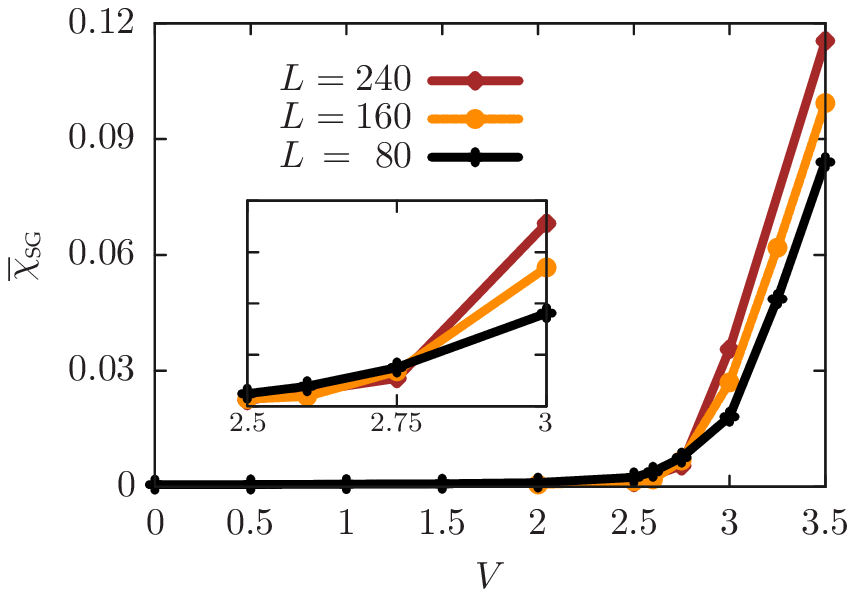}
\vskip-8pt
\caption{
Edwards-Anderson order parameter (top) and condensate fraction $\rho_c/\rho$ (middle) as functions of $V$ at half-filling,
computed via the cavity method at different values of $\beta$. In the middle panel $\rho_c/\rho$ as obtained by
SGF at $\beta=5$ is reported.
(Bottom) Scaled spin-glass susceptibility $\overline{\chi}_{\text{\tiny{SG}}} = \chi_{\text{\tiny{SG}}}/L^{5/6}$ reported as a function of
$V$; standard finite-size-scaling arguments~\cite{Guo:1994sf} show that the different curves must intersect at the spin-glass
transition. \label{fig:Half-fill}
}

\end{figure}
\paragraph*{Model.}

Strongly interacting bosons on a lattice can be conveniently described
by means of the extended Hubbard Hamiltonian, namely 
\begin{eqnarray}\label{eq:Hdef}
\widehat{\mathcal{H}} & = & -t\sum_{\left\langle i,j\right\rangle }\left[{b}_{i}^{\dagger}{b}_{j}+{b}_{i}{b}_{j}^{\dagger}\right]+V\sum_{\left\langle i,j\right\rangle }{n}_{i}{n}_{j},\label{eq:ExtHam}
\end{eqnarray}
where ${b}_{i}^{\dagger}$(${b}_{i}$) creates (destroys) a hard-core
boson on site $i$, ${n}_{i}={b}_{i}^{\dagger}{b}_{i}$ is the site-density
and the summations over the indexes $\left\langle i,j\right\rangle $
are extended to nearest-neighbouring vertices of a given lattice with
$L$ sites. In the following we will set $t=1$, i.e. we will measure all energies in units of $t$.
In this work, to capture the essential physics of the problem in exam, 
we adopt a minimal and transparent strategy to induce
geometrical frustration in the solid. We therefore consider the set of {\it all possible
graphs} of $L$ sites, such that each site is connected to {\it exactly} $z=3$ other sites, 
and give the same probability to each graph in this set.
We will discuss average properties over this {\it ensemble}
of random graphs in the thermodynamic limit $L\to\infty$.
The motivations for this choice are the following:
{\it i)}~On a square lattice, model (\ref{eq:Hdef}) is known to produce a solid insulating 
phase at high enough density, where the particles are arranged in a
checkerboard pattern \cite{PhysRevLett.94.207202}. This is due to the fact that all
loops have even length. On the contrary, typical random graphs
are characterized by loops of even or odd length;
in the classical case $t=0$, this frustrates the solid phase enough to produce a 
{\it thermodynamically stable} glass phase at high density~\cite{mezard2001bethe}. 
{\it ii)}~Typical random graphs have the important property that they are locally
isomorphic to trees, since the size of the loops scales as $\ln L$ for large $L$: 
indeed, this is a consistent way of defining Bethe lattices 
without boundary~\cite{mezard2001bethe}. This locally tree-like structure
allows to solve the model exactly, at least in the liquid phase,
by means of the cavity method \cite{laumann:134424,semerjian:014524}.
{\it iii)}~These lattices are quite different from square lattices. Yet, it has been shown 
in the classical case, and for some more complicated interactions, that the phase 
diagram is qualitatively very similar for the model defined on a random graph and
on a square lattice~\cite{PhysRevLett.88.025501,PhysRevE.67.057105}.
Hence, we believe that it is possible to find a model similar to Eq.~(\ref{eq:Hdef}), defined
on a square lattice but with slightly more complicated interactions (probably involving many-body
terms) that will show the same qualitative behaviour of the model investigated here.

\paragraph*{Methods.}

The stochastic sampling of the quantum partition function $\mathcal{Z} = \text{Tr}\, e^{- \beta \widehat{\mathcal{H}} }$ 
at finite temperature $T=1/\beta$ can be conveniently exploited to obtain numerically exact 
properties of a generic bosonic Hamiltonian such as \eqref{eq:ExtHam}.
Quantum Monte Carlo (QMC) schemes based on the original Worm algorithm idea~\cite{worm} have been recently extended to Canonical ensemble simulations~\cite{Rombouts:2006wd, Rousseau:2008hl}.
These methods offer an efficient scheme based on the sampling of the configuration space spanned by the extended partition 
function $\mathcal{Z}_w(\tau) = \text{Tr}\, e^{-(\beta - \tau) \widehat{\mathcal{H}} }\widehat{W} e^{- \tau \widehat{\mathcal{H}} }$,
where $\widehat{W}$ is a suitable \textit{worm} operator determining an imaginary time discontinuity in the world-lines.
We have chosen the worm operator introduced in \cite{Rousseau:2008hl}, which 
is a linear superposition of $n$-body Green functions, avoiding the complications
arising in~\cite{Rombouts:2006wd} where the commutability of the worm operator with the non-diagonal part 
of the Hamiltonian is required. Full details of the Stochastic Green Function (SGF) method are described in 
Ref.~\cite{Rousseau:2008hl}, we only stress here that access to exact equal-time 
thermal averages of $n$-body Green functions is granted as well as to thermal averages of imaginary time correlation 
functions of local, i.e. diagonal in the occupation numbers representation, quantum operators.

A different and complementary approach to models defined on random lattices
consists in solving them exactly
in the thermodynamic limit $L \to \infty$, by means of the {\it cavity method}~\cite{mezard2001bethe}. 
Since local observables are self-averaging in
this limit, this results in automatically taking into account the average over the different realisations of the random graphs.
For bosonic systems, 
the cavity method allows to reduce the solution of the model
to the problem of finding the fixed point of a functional equation
for the local effective action, in a similar spirit to bosonic DMFT.
All the details of the computation have been discussed
in~\cite{semerjian:014524}, where it has been shown that the method allows to compute the average
of all the relevant observables.
However, in the simplest version discussed in~\cite{semerjian:014524}, the cavity method can only describe
homogeneous pure phases such as the low-density liquid. In order to describe exactly 
the high density glassy phase, where many different inhomogeneous states coexist,
one has to introduce a generalization of the simplest cavity method which goes under
the name of {\it replica symmetry breaking} (RSB). Unfortunately, this is already a difficult
task for classical models, in particular in spin-glass like phases~\cite{mezard2001bethe}. Hence, in this paper
we describe the glassy phase using the simplest version of the method, the so-called
replica symmetric (RS) one. This yields an approximate description of the glassy phase which
we expect to be qualitatively correct.
To summarize, in the low-density liquid phase we can compute averages numerically
with SGF and analytically with the cavity method, and we obtain a perfect agreement between
the two results. In the glassy phase, the RS cavity method is only approximate, an exact solution for $L\to\infty$ requiring the introduction
of RSB. On the other hand, SGF is limited for large $L$ by the unavoidable divergence 
of equilibration times due to the glassy nature of the system.
Still, we find a good agreement between the result of SGF for fairly large $L$, where the system can still be equilibrated, 
and the RS cavity method for $L\to \infty$, 
making us confident that the qualitative and quantitative picture of the glassy phase we
obtained here is fully consistent. Moreover, we solved the model at the simplest (one-step) RSB level
in some selected state points and we found a very small quantitative difference with the RS solution.

\begin{figure}
\includegraphics[width=7.5 cm]{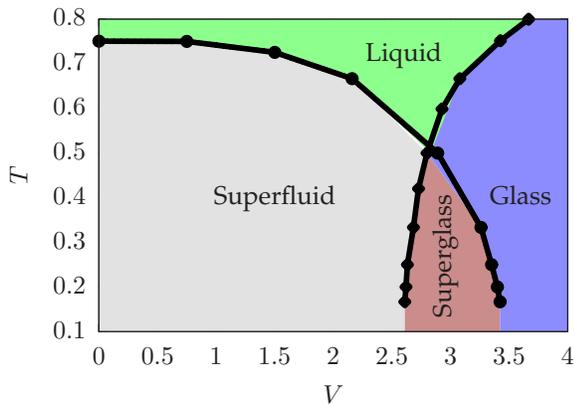}
\caption{
Finite temperature phase diagram at half-filling. \label{fig:pd}
}
\vskip-8pt
\end{figure}

\paragraph*{Results.}
The presence of off-diagonal long range order can be conveniently detected by 
considering the large separation limit of the one-body density matrix, 
i.e. the condensate density reads 
\begin{eqnarray}
\rho_{c} & = & \lim_{\left|i-j\right|\rightarrow\infty}\overline{\left\langle b_{i}^{\dagger}b_{j}\right\rangle }
= \overline{|\langle b_{i} \rangle^2|} \ ,
\label{eq:rhocond}\end{eqnarray}
where the square brackets indicate a quantum and thermal average and the bar
indicates averages over independent realizations of the random graphs.
The cavity method works in the grand-canonical ensemble and gives direct access
to the average of $b$, while canonical ensemble simulations done with SGF give easy access to the one-body density matrix.
On the other hand, spin-glass order is signaled by the breaking of translational invariance, namely 
$\langle n_i \rangle \neq L^{-1} \sum_{i=1}^L \langle n_i \rangle = \rho$.
Introducing $\delta n_{i}=\left(n_{i}-\rho\right)$, the on-site deviation
from the average density,
spin-glass order can be quantified by the
Edwards-Anderson order parameter
\begin{eqnarray}
q_{\text{\tiny{EA}}} = \frac{1}{L} \sum_{i=1}^L \overline{\langle \delta n_i \rangle^2} \ ,
\end{eqnarray}
which can be easily computed by the cavity method,
or by the divergence of the spin-glass susceptibility 
\begin{eqnarray}
\chi_{\text{\tiny{SG}}} & = & \frac{1}{L} \int^\beta_0 d\tau \sum_{i,j}\overline{\left\langle \delta n_{i}(0)\delta n_{j}(\tau)\right\rangle ^{2}} \ ,
\label{eq:chisg}\end{eqnarray}
which is more easily accessible in SGF.
It is possible to show \cite{Guo:1994sf} that $\chi_{\text{\tiny{SG}}}$ is the susceptibility naturally associated to the order
parameter $q_{\text{\tiny{EA}}}$, because it can be defined as the derivative of $q_{\text{\tiny{EA}}}$ with respect to an external field
coupled to the order parameter itself (as in standard critical phenomena).

At half-filling factor $\rho=1/2$, the condensate fraction, the Edwards-Anderson order parameter, and the 
scaled spin-glass susceptibility are shown in Fig.~\ref{fig:Half-fill}.
In middle panel we compare the values of condensate fraction obtained via the cavity method and via SGF in a linear
extrapolation to $L\to\infty$. The very good coincidence of these results supports our conjecture
that the approximate RS description of the glass phase we adopted here 
is quantitatively and qualitatively accurate.
At the lowest temperature, the system becomes a glass around 
$V \sim 2.7$ while it still displays BEC; the condensate
fraction only vanishes at $V\sim 3.5$ inside the glass phase.
This clearly establishes the existence of a zero-temperature superglass phase in the region
$2.7 \lesssim V \lesssim 3.5$. Note additionally that both
transitions are of second order, hence the condensate fraction is
a continuous function; since the latter stays finite on approaching the spin-glass 
transition from the liquid side (where the cavity method gives the exact solution),
it must also be finite on the glass side just after the transition.
In Fig.~\ref{fig:pd} we report the finite temperature phase diagram of the model at half-filling.
It is defined by two lines: the first separates the non-condensed ($\langle b \rangle$=0) from
the BEC ($\langle b \rangle\neq 0$) phase, the second separates the glassy ($q_{\text{\tiny{EA}}}\neq 0$) from
the liquid ($q_{\text{\tiny{EA}}}=0$) phase. The intersection between these two lines determines the existence of
four different phases (normal liquid, superfluid, normal glass, superglass).

\paragraph*{Ground-state degeneracy.}

\begin{figure}
\includegraphics[width=7.5 cm]{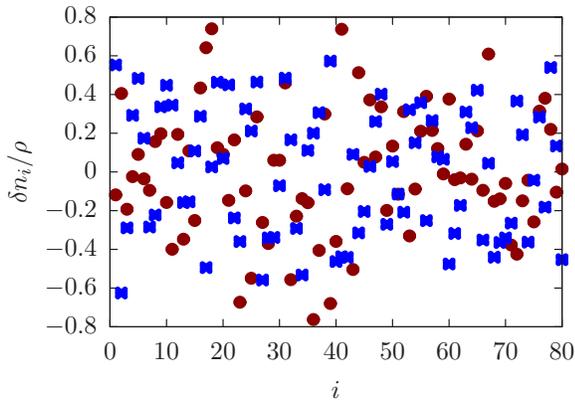}
\vskip-10pt
\caption{Variational expectation values of the site density for different sets
of the optimized parameters at half-filling density for $L=80$ and
$V=4$.\textcolor{red}{ \label{fig:Degener}}}
\end{figure}

Geometrical frustration induces the existence of a highly degenerate
set of ground-states, each of them characterized by a different
average on-site density, 
which is absent in glassy phases induced
by localization in disordered external potentials such as the Bose glass.
To demonstrate this peculiar feature, it
is instructive to consider a variational wave-function explicitly
breaking the translational symmetry of the lattice 
\begin{eqnarray}
\left\langle \mathbf{n}\right.\left|\Psi_{\alpha}\right\rangle  & \propto & \exp\left[\sum_{i}\alpha_{i}n_{i}\right],\label{eq:psialpha}
\end{eqnarray} where the variational parameters $\alpha_{i}$ are explicitly site-dependent
and tend to (dis)-favour the occupation of a given site. In the spin-glass
phase of the bosons, the optimal set of the variational parameters
is highly dependent on the initial conditions associated with the
$\alpha_{i}$, whereas all the variational states, even with different
parameters, have almost degenerate variational energies. Each set
of optimized variational parameters is then representative of one
of the many degenerate ground-states of $\widehat{\mathcal{H}}$. As an example,
we show in Fig.~\ref{fig:Degener} the variational expectation values
of the site densities for two different solutions resulting from the minimization of the variational energy 
with the SRH method \citep{Sorella:2005fk}, a robust stochastic variant of the Newton Method. 
We further checked, using the zero-temperature Green Function Monte Carlo method \citep{SorellaCapriotti2000}, 
that if one applies the imaginary-time evolution $| \phi_\alpha \rangle = \exp(-\tau \widehat{\mathcal{H}}) | \Psi_\alpha \rangle$
to one of these states, the density profile remains amorphous for a time $\tau$ that
is divergent with the size of the system.

\paragraph*{Conclusions.}

The aim of this Letter was to show the existence of a stable superglass phase in a lattice model of geometrically 
frustrated bosons, in absence of quenched disorder in the Hamiltonian. This has been done by combining the analytical solution of the model via the quantum cavity method
and numerical simulations via Quantum Monte Carlo.
The glass phase we found is very different from the usual Bose glass, since the latter is driven by localization
effects in presence of an external disorder and is then insulating, while the former
is driven by self-induced frustration on a disordered lattice and displays Bose-Einstein condensation. This results in a coexistence
of a large number of degenerate amorphous ground states, whose existence we showed by a variational argument
corroborated by QMC.
We expect, by analogy with the classical case~\cite{PhysRevLett.88.025501}, 
that the glassy phase found here will 
exist also on regular finite dimensional lattices, provided the interactions are modified to induce sufficient geometrical frustration. 
In that case, its properties should be very
similar to the one showed by metastable superglasses observed both in numerical simulations~\citep{boninsegni:105301} 
and experiments~\cite{B.Hunt05012009} on Helium 4. The main difference is that, due to the randomness of
the underlying lattice,
the superglass studied here is a {\it truly stable equilibrium state}, 
allowing for a much more precise characterization of its properties.

\paragraph*{Acknowledgements.}

We wish to warmly thank G.~Semerjian who participated to an early stage of this work and gave us many
important suggestions. We also thank S.~Baroni, F.~Becca, G.~Biroli, S.~Moroni, and S.~Sorella 
for many precious discussions. G.C. acknowledges the allocation of computer resources at
the CINECA supercomputing center from the CNR-INFM \textit{Iniziativa
Calcolo per la Fisica della Materia} program. 

\paragraph*{Note added.}
After this work was completed, we became aware of the paper~\cite{arXiv:0909.1845} where related results have been obtained.

\bibliographystyle{apsrev}

\end{document}